\documentclass[sigconf]{acmart}

\usepackage{pifont}
\usepackage{bm}
\usepackage{algorithm}
\usepackage{algorithmic}
\usepackage{comment}
\usepackage{multirow}
\usepackage{booktabs}
\usepackage{stmaryrd}
\usepackage{enumerate}

\newtheorem{myDef}{Definition}

\AtBeginDocument{%
  \providecommand\BibTeX{{%
    \normalfont B\kern-0.5em{\scshape i\kern-0.25em b}\kern-0.8em\TeX}}}

\setcopyright{iw3c2w3}
\copyrightyear{2021}
\acmYear{2021}
\acmDOI{10.1145/3442381.3449803}

\acmConference[WWW '21]{Proceedings of the Web Conference 2021}{April 19--23, 2021}{Ljubljana, Slovenia}
\acmBooktitle{Proceedings of the Web Conference 2021 (WWW '21), April 19--23, 2021,
Ljubljana, Slovenia}
\acmPrice{}
\acmISBN{978-1-4503-8312-7/21/04}
\setcopyright{acmcopyright}
\begin{document}

\title{Itinerary-aware Personalized Deep Matching at Fliggy}


\author{Jia Xu}
\email{xujia@gxu.edu.cn}
\orcid{0000-0003-4061-8262}
%\email{webmaster@marysville-ohio.com}
\affiliation{%
	\institution{College of Computer, Electronics and Information, Guangxi University}
	%\streetaddress{P.O. Box 1212}
	\city{Nanning}
	%\state{Guangxi}
	\country{China}
	\postcode{530004}
}

\author{Ziyi Wang}
\authornote{Ziyi Wang is the corresponding author.}
%\author{Zulong Chen}
%\email{zulong.cz@alibaba-inc.com}
\affiliation{
	\institution{Alibaba Group}
	\city{Hangzhou}
	\country{China}}
\email{jianghu.wzy@alibaba-inc.com}

\author{Zulong Chen}
\affiliation{
	\institution{Alibaba Group}
	\city{Hangzhou}
	\country{China}}
\email{zulong.cz@alibaba-inc.com}

\author{Detao Lv}
\affiliation{
	\institution{Alibaba Group}
	\city{Hangzhou}
	\country{China}}
\email{detao.ldt@alibaba-inc.com}

\author{Yao Yu}
\affiliation{
	\institution{Alibaba Group}
	\city{Hangzhou}
	\country{China}}
\email{sichen.yy@alibaba-inc.com}

\author{Chuanfei Xu}
\affiliation{
	\institution{Concordia University}
	\city{Montreal}
	\state{Quebec}
	\country{Canada}}
\email{chuanfeixu@126.com}

\renewcommand{\shortauthors}{Xu, et al.}

\begin{abstract}
Matching items for a user from a travel item pool of large cardinality have been the most important technology for increasing the business at Fliggy, one of the most popular online travel platforms (OTPs) in China. There are three major challenges facing OTPs: sparsity, diversity, and implicitness. In this paper, we present a novel Fliggy ITinerary-aware deep matching NETwork (\textbf{FitNET}) to address these three challenges. FitNET is designed based on the popular deep matching network, which has been successfully employed in many industrial recommendation systems, due to its effectiveness. The concept \textit{itinerary} is firstly proposed under the context of recommendation systems for OTPs, which is defined as the list of unconsumed orders of a user. All orders in a user itinerary are learned as a whole, based on which the implicit travel intention of each user can be more accurately inferred. To alleviate the sparsity problem, users' profiles are incorporated into FitNET. Meanwhile, a series of itinerary-aware attention mechanisms that capture the vital interactions between user's itinerary and other input categories are carefully designed. These mechanisms are very helpful in inferring a user's travel intention or preference, and handling the diversity in a user's need. Further, two training objectives, i.e., prediction accuracy of user's travel intention and prediction accuracy of user's click behavior, are utilized by FitNET, so that these two objectives can be optimized simultaneously. An offline experiment on Fliggy production dataset with over 0.27 million users and 1.55 million travel items, and an online A/B test both show that \textbf{FitNET} effectively learns users' travel intentions, preferences, and diverse needs, based on their itineraries and gains superior performance compared with state-of-the-art methods. \textbf{FitNET} now has been successfully deployed at Fliggy, serving major online traffic. 
\end{abstract}




\begin{comment}
\begin{teaserfigure}
  \includegraphics[width=\textwidth]{sampleteaser}
  \caption{Seattle Mariners at Spring Training, 2010.}
  \Description{Enjoying the baseball game from the third-base
  seats. Ichiro Suzuki preparing to bat.}
  \label{fig:teaser}
\end{teaserfigure}
\end{comment}


\copyrightyear{2021}
\acmYear{2021} 
\acmConference[WWW '21]{Proceedings of the Web Conference 2021}{April 19--23,
	2021}{Ljubljana, Slovenia}
\acmBooktitle{Proceedings of the Web Conference 2021 (WWW '21), April 19--23, 2021,
	Ljubljana, Slovenia}
\acmPrice{}
\acmPrice{}
\acmDOI{10.1145/3442381.3449803}
\acmISBN{978-1-4503-8312-7/21/04}

\maketitle

\section{Introduction}
\label{sec:intro}
\begin{figure}[h]
	\centering
	\includegraphics[width=\linewidth]{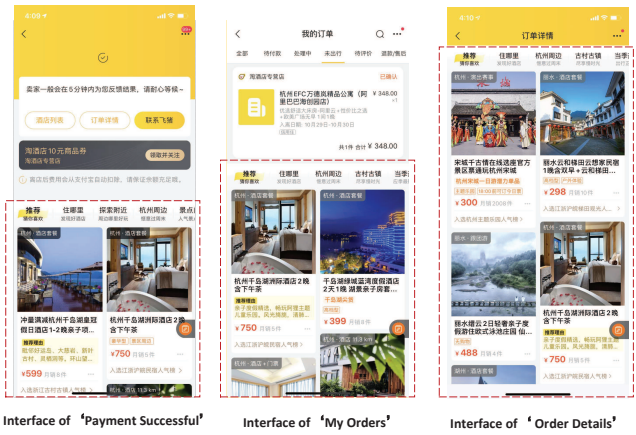}
	\caption{Areas highlighted with dashed rectangles are personalized for tens millions of users in Mobile Fliggy App.}
	\label{fig:interface}
\end{figure}

Fliggy from Alibaba group\footnote{\url{https://www.fliggy.com/}}, is one of the most popular online travel platforms (OTPs) in China, which serves tens of millions of online users by providing million-scale travel-related items (such as air ticket, hotel, and package tour). With the continuous growth in the amount of users and items at Fliggy, it is increasingly critical to help users find items that they might need and prefer to. Therefore,  
in recent years, great efforts have been paid to develop the Fliggy recommendation system (FRS), which has made significant contribution to the optimization of user experience and the elevation of business value. For example, several important interfaces on Mobile Fliggy App (see Figure \ref{fig:interface}) use FRS to automatically generate personalized items to satisfy users' needs, and serve major traffic. 

\begin{figure}[h]
	\centering
	\includegraphics[width= 160 pt]{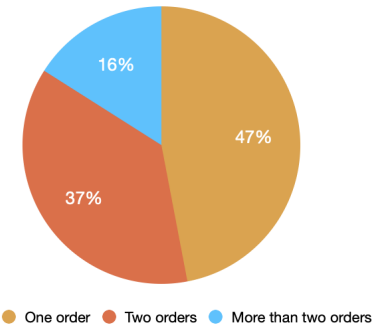}
	\caption{Statistics of orders created by Fliggy users in terms of a travel plan (September 10, 2020 to October 10, 2020).}
	\label{fig:order}
\end{figure}

\begin{figure}[h]
	\centering
	\includegraphics[width= 240 pt]{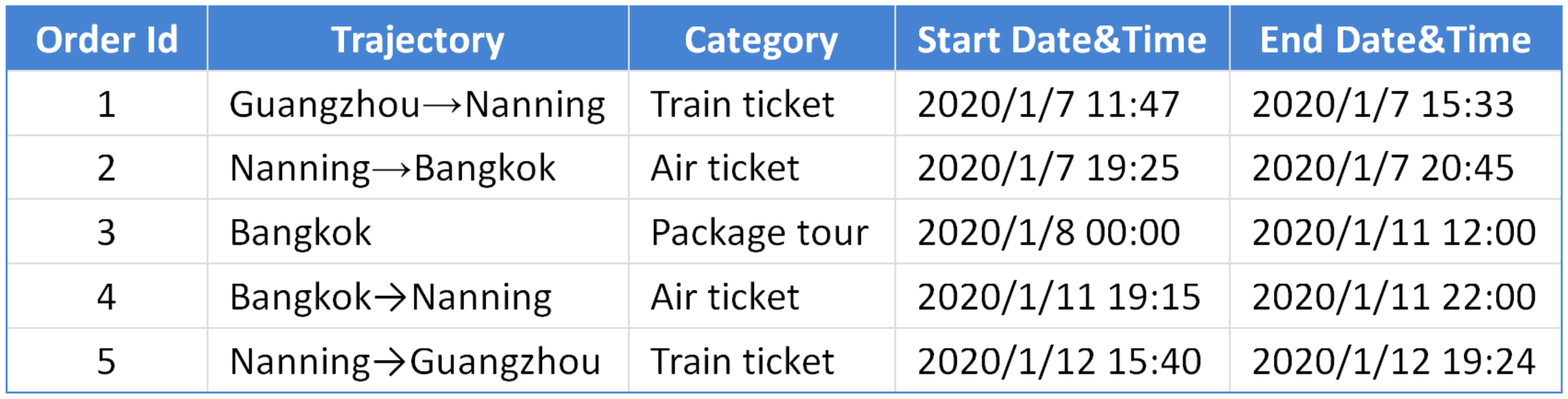}
	\caption{A user's orders created at Fliggy for a travel.}
	\label{fig:1}
\end{figure}

With large-scale users and items, the recommendation task in FRS follows the popular two-phase model, which is composed of the matching phase and the ranking phase, used by many mainstream industrial recommendation systems \cite{YouTubeDNN16, MIND19}, to make sure the recommended items for users are generated in an efficient way. In the matching phase, thousands of candidate items are retrieved from an item pool of million-scale items for each user. Then, the ranking phase is responsible for predicting probabilities of users interacting with these candidate items, and finally generates recommendations. The quality of candidate items retrieved in the matching phase plays a key role in the whole model \cite{SDM19}. Currently, inspired by the success of deep learning from computer vision domain \cite{Huang_2017_CVPR} and natural language processing (NLP) \cite{2014Neural}, many deep matching networks (DMNs) that match items for users on the basis of deep neural networks have been successfully deployed in many industrial recommendation systems, due to their effectiveness and efficiency \cite{YouTubeDNN16, DeepWalk+18, ENCORE18, SDM19, MIND19}. However, without considering the challenges determined by business characteristics of OTPs, these networks become ineffective when the matching objects are travel-related items, which makes the personalized deep matching problem at Fliggy extremely challenging. Major challenges facing OTPs are: 
\begin{itemize}  
	\item \textbf{Sparsity.} Compared with other e-commerce platforms (e.g., Taobao in China), users' behaviors collected at OTPs are more sparse, for travel is usually an infrequent need of users.
	\item \textbf{Diversity.} When a user plans a travel, the user may have cross-domain diverse needs, including buying air ticket, booking hotel, calling taxi service, choosing a package tour, etc. The diversity of a user's needs must be considered in recommendation process.  
	\item \textbf{Implicitness.} According to statistics reported by Fliggy from Aug. 10 to Sep. 10 of the year 2020, over $53\%$ users create more than one order for a travel, and $16\%$ of users even have more than two orders (see Figure \ref{fig:order} for details). The reported results reveal that a user's travel intention, which is of importance in matching items for the user, is generally implicitly delivered by multiple orders rather a single order. For example, Figure \ref{fig:1} displays a user's orders created at Fliggy for a travel. By considering those orders as a whole, we can accurately infer the user's travel intention of the third order is for sightseeing, while the intention of the remaining orders is for transfer. Under such circumstance, tourist items of Nanning (i.e., the transit city) should not be presented to the user.    
\end{itemize}

In this paper, we focus on the above challenges in the matching phase of OTPs and address the above challenges through the creation of a novel \underline{F}liggy \underline{IT}inerary-aware deep matching \underline{NET}work called \underline{FitNET}. On one hand, to cope with the \textbf{implicitness} problem, the proposed network uses all unconsumed orders of a user which are deemed to correspond to a travel as a whole input category, defined as \textit{itinerary}, to enhance the understanding of user's intention in each unconsumed order. 
Besides user's itinerary information, user's profiles which consists of abundant user's side information are incorporated into the framework FitNET, so as to alleviate the \textbf{sparsity} and cold start problems.
On the other hand, inspired by the recent successes of attention mechanism \cite{10.5555/3295222.3295349, DBLP:conf/icdm/KangM18, zhang2019next}, a series of itinerary-aware attention mechanisms are carefully designed in FitNET to capture abundant interactions between the embedding vectors of users' itineraries and the embedding vectors of other input categories. The proposed attention mechanisms offer great help in improving the accuracy of inferring users' travel intentions, identifying users' preferences to an item's features (such as destination city and category), and understanding users' \textbf{diverse} needs.
Moreover, the network loss is defined by considering both of the prediction accuracy to users' travel intentions (e.g., sightseeing or others) and the prediction accuracy towards the click behavior of a user to an impression item. These two aspects of loss are seamlessly integrated together under the FitNET framework, so that they can get feedback from each other. The main contributions of this work are summarized below:
\begin{itemize}  
	\item To the best of our knowledge, this is the first work that proposes to leverage a user's \textit{itinerary} (i.e., the list of unconsumed orders of the user) rather than a single order of the user to optimize the deep matching phase of a large-scale recommendation system at OTPs. 
	
	\item We design a novel itinerary-aware deep matching network (FitNET for short) for Fliggy, one of the most popular online travel platforms (OTPs) in China, to improve the effectiveness of deep neural network in matching personalized travel-related items from a item pool with large cardinality.
	In FitNET, unconsumed orders in a user's itinerary is considered as a whole, and a series of carefully designed itinerary-aware attention mechanisms are proposed, which enables FitNET to cope with the three tough challenges facing OTPs effectively. 
	
	\item We verify FitNET on Fliggy production dataset
	with over $0.27$ million users and $1.55$ million travel items. 
	Experimental results show that FitNET effectively learns users' travel intentions and preferences based on their itinerary information and beats state-of-the-art methods. In particular, FitNET improves the average CTR by $1.3\%$ compared with the next best method in an online A/B test, which is an apparent progress made by an industrial recommendation system. Recently, FitNET has been deployed at Fliggy successfully, serving most online traffic.  
\end{itemize}

The remainder of this paper is organized as follows: related works are reviewed in Section \ref{sec:related}; Section \ref{sec:fitnet} elaborates the technical details of the FitNET model; In Section \ref{sec:exp}, we detail the experiments for comparing FitNET with existing methods; The last section gives conclusions and points out future directions of this paper.

\section{Related Works}
\label{sec:related}
Most personalized recommendation methods are designed based on the intuition that users’ interests can be inferred from their historical behaviors or from other users with similar preferences \cite{tree18}. Based on such intuition, enormous personalized recommendation systems (RSs) have been proposed and deployed from all walks of life, which consequently makes people's life more convenient. 

To further improve users' click-through rate (CTR), post-click conversion rate (CVR), and time spent in applications, collaborative filtering (CF), which makes use of user-item interactions (e.g., clicks) to learn the relevance between users and items, is proposed and soon becomes the key technique for RSs \cite{33, 55, 66, 99, www14, 77_18}. 
In particular, the item-to-item (i2i) recommendation, which recommends items similar to an item of interest in terms of different similarity measures, has been the most widely used CF method under the industrial settings \cite{44,55,DBLP:conf/kdd/KabburNK13}, owing to its interpretability and efficiency in real-time personalization \cite{22}.
Being different to the i2i method, the item-relationship CF recommendation is another popular research domain. It focuses on discovering relationships among items -- such as substitutes or complements -- that go beyond traditional item similarities \cite{itemRelation15}. 

Generally, a recommendation procedure consists of two phases: matching and ranking. 
In the matching phase, thousands of candidate items are retrieved by matching models from an item pool with large cardinality. During the ranking phase, a more complex ranking model is used to assign comparable scores to the candidate items, and top-$k$ items are recommended to users.
After the big success achieved in computer vision domain \cite{imageNET12, DBLP:journals/corr/SimonyanZ14a, Huang_2017_CVPR}, the deep neural network \cite{Hinton06} has become the most popular technique in both academia and industry circles for solving the matching and ranking problems in RSs \cite{YouTubeDNN16, DIN17, DeepWalk+18, Airbnb18, ENCORE18, MIND19, SDM19, Fliggy20}, due to its better representation and generalization ability compared with CF based solutions \cite{SDM19}. 

Since the quality of retrieved items in the matching phase directly determines the upper quality limit of the ranking phase, matching model plays a vital role in the whole recommendation procedure. Recent years witness a good many works that optimize the matching phase by use of the deep neural network (called as the deep matching network, DMN for short) and achieves remarkable effectiveness.
For example, a deep neural network architecture was designed for both of the matching phase and the ranking phase in YouTube recommendation, named YouTube DNN \cite{YouTubeDNN16}. In YouTube DNN, items and user are mapped to vectors via embedding layers and a feedforward neural network. At serving time, Youtube DNN computes the most likely $k$ videos and presents them to the user.
Multi-interest network with dynamic routing (MIND) \cite{MIND19} then was proposed to deal with users' diverse interests in the matching stage. In MIND, a multi-interest extractor layer based on capsule routing mechanism is designed, which is applicable for clustering users' historical behaviors and extracting users' diverse interests. 
In \cite{DeepWalk15}, Bryan et. al proposed the DeepWalk model, a novel method which employs local information obtained from truncated random walk to learn latent representations of vertices in a network. After that, Wang et. al. proposed to apply the DeepWalk as a DMN to improve the relevance of matching results in the matching phase \cite{DeepWalk+18}. In specific, they construct an item graph based on users’ historical behaviors, and learns the embeddings of all items in the graph. 
At the same time, Zhang et. al. presented the ENCORE model \cite{ENCORE18}, which is a neural item-relationship based recommender focusing on matching and presenting "complementary" items to users. 
Then, a sequential deep matching model, SDM \cite{SDM19}, was proposed to capture users’ dynamic preferences in the matching phase by combining users' short-term sessions and long-term behaviors.  
Currently, the idea of collaborative filtering is also connected with the deep neural network. For example, 
collaborative filtering is formulated as a deep neural network in \cite{DBLP:conf/kdd/WangWY15}, which jointly performs deep representation learning for the content information and collaborative filtering for the ratings (i.e., feedback) matrix. 
Moreover, a novel CF model, AutoRec, was proposed based on the autoencoder paradigm \cite{CF+DL15}. And a neural network based collaborative filtering method, NCF, was presented in \cite{CF+DL17}, which complements the mainstream shallow models for collaborative filtering, opening up a new avenue of research possibilities for recommendation based on deep learning.

Inspired by the big success of deep learning network in the circle of recommendation, our FitNET model follows the deep neural network, and focuses on optimizing the matching phase for online travel platforms (OTPs). 
Although previous DMN-based works provide many good ideas in improving the matching phase, these works can not well address the challenges faced by online travel applications, namely, sparsity, diversity, and implicitness. In this work, we consider these challenges by introducing the concept of `\textit{itinerary}' and carefully design a series of effective attention mechanisms by being aware of the information of user itinerary. In the following, we present the design of our FitNET model.

\section{FitNET MODEL}
\label{sec:fitnet}
In this section, we first formalize the personalized matching problem faced by Fliggy Recommendation System (FRS), and then present the FitNET model which effectively solves the problem at Fliggy. 

\subsection{Problem Formalization}
\label{sec:problem}
Let $\mathcal{U}$ be the set of users at Fliggy. The objective of personalized matching in FRS is to improve the quality of personalized item set retrieved from an item pool $\mathcal{P}$ of million-scale items related to travel matters for each user  $u_i\in \mathcal{U}$, by addressing before-mentioned three challenges facing OTPs: namely, sparsity, diversity, and implicitness.
It is known that the quality of the retrieved personalized item set is highly influenced by the inferred travel intention of the user. 

To achieve the objective of personalized matching in FRS, this paper designs a novel model named Fliggy itinerary-aware deep matching network (FitNET), which overcomes the challenges facing OTPs by introducing users' itinerary information into the neural network. The concept of \textit{itinerary} is formally defined as

\begin{myDef}
	\label{DEF-Itinerary}
	\textbf{\textit{Itinerary}}.  For a given user $u_i\in \mathcal{U}$, let $\bar{o}_{u_i}$ denotes an order of $u_i$ at Fliggy which has not yet been consumed by $u_i$. The itinerary of $u$, denoted by $\mathcal{I}_{u_i}$, is the list of unconsumed orders of $u_i$ at Fliggy (i.e., $\mathcal{I}_{u_i}=\{\bar{o}_{u_i}\}$) which is deemed to correspond to a travel of the user. 
\end{myDef}

The implicitness problem of OTPs tells that it is quite difficult to infer a user's travel intention with a single order of the user. The introduction of 'itinerary' enables FitNET to be trained based on the granularity of a user's itinerary rather than the granularity of the user's single order. As a result, FitNET can make better prediction to the travel intention of each user, which tackles the implicitness challenge facing OTPs.   

Besides the itinerary of each user (i.e., $\mathcal{I}_{u_i}, {u_i}\in \mathcal{U}$), basic profile of every user is also used as the input, so as to lighten the sparsity challenge due to the inadequate amount of user behaviors collected at OTPs. Let  $\mathcal{P}_{u_i}$ denote the basic profile of user $u_i$ (e.g., permanently reside city id, home city id, and etc.), $\mathcal{B}_{u_i}$ be a sequence of items interacted by user $u_i$ before the first order in $\mathcal{I}_{u_i}$ is created by $u_i$ (also known as the user behavior sequence), and $\mathcal{F}_{u_i}$ represent the target item features (e.g., item id, category id, and etc.). Then, every input instance of historical data in FitNET can be represented by a quadruple $(\mathcal{P}_{u_i}, \mathcal{I}_{u_i}, \mathcal{B}_{u_i}, \mathcal{F}_{u_i})$. 

The critical task of FitNET is to learn a function for mapping user-related raw inputs into user representations, which can be formulated by

\begin{equation}
\label{eq:1}
\bm{v}_{u_i} = f_{\mathcal{U}}(\bm{e}(\mathcal{P}_{u_i}), \bm{e}(\mathcal{I}_{u_i}), \bm{e}(\mathcal{B}_{u_i})),
\end{equation}
where, $v_{u_i}$ is the representation vector of user $u_i$ computed based on the embedding vectors of $\mathcal{P}_{u_i}$, $\mathcal{I}_{u_i}$, and $\mathcal{B}_{u_i}$, represented as $\bm{e}(\mathcal{P}_{u_i})$, $\bm{e}(\mathcal{I}_{u_i})$, and $\bm{e}(\mathcal{B}_{u_i})$, respectively. Besides, let $\mathcal{T}$ be the set of target items, and $v_{t_j}$ be the representation vector of a target item $t_j \in \mathcal{T}$ . $v_{t_j}$ can be obtained by the learned mapping function $f_\mathcal{T}$ of FitNET with the embedding vector of target item features $\mathcal{F}_{t_j}$, denoted as $\bm{e}(\mathcal{F}_{t_j})$), as the input (see Equation \ref{eq:2} for details). Specific implementation of the embedding strategy will be elaborated in Section \ref{sec:embed}.

\begin{equation}
\label{eq:2}
\bm{v}_{t_j} = f_{\mathcal{T}}(\bm{e}(\mathcal{F}_{t_j}))
\end{equation}

In FitNET, the representation vectors for both users and items are optimized by the employment of a group of itinerary-aware attention mechanisms, which leverage users' itinerary information to improve the embeddings of intermediate results. These attention mechanisms offer great help not only to the inference of users' travel intentions or preferences, but also to the identification of a user's diverse needs, which addresses the diversity challenge faced by OTPs.  
When the representation vector of a user $u_i \in \mathcal{U}$ and the representation vectors of items $t_j \in \mathcal{P}$ are learned by FitNET, the top $k$ items in $\mathcal{P}$ are presented as the matched results for the user $u_i$, according to the score function:  

\begin{equation}
\label{eq:3}
f_{score}(\bm{v}_{u_i}, \bm{v}_{t_j}) = \bm{v}_{u_i} \cdot \bm{v}_{t_j}.
\end{equation}
where the operator $\cdot$ represents the inner product of two vectors.

\begin{figure*}[h]
	\centering
	\includegraphics[width= 450 pt]{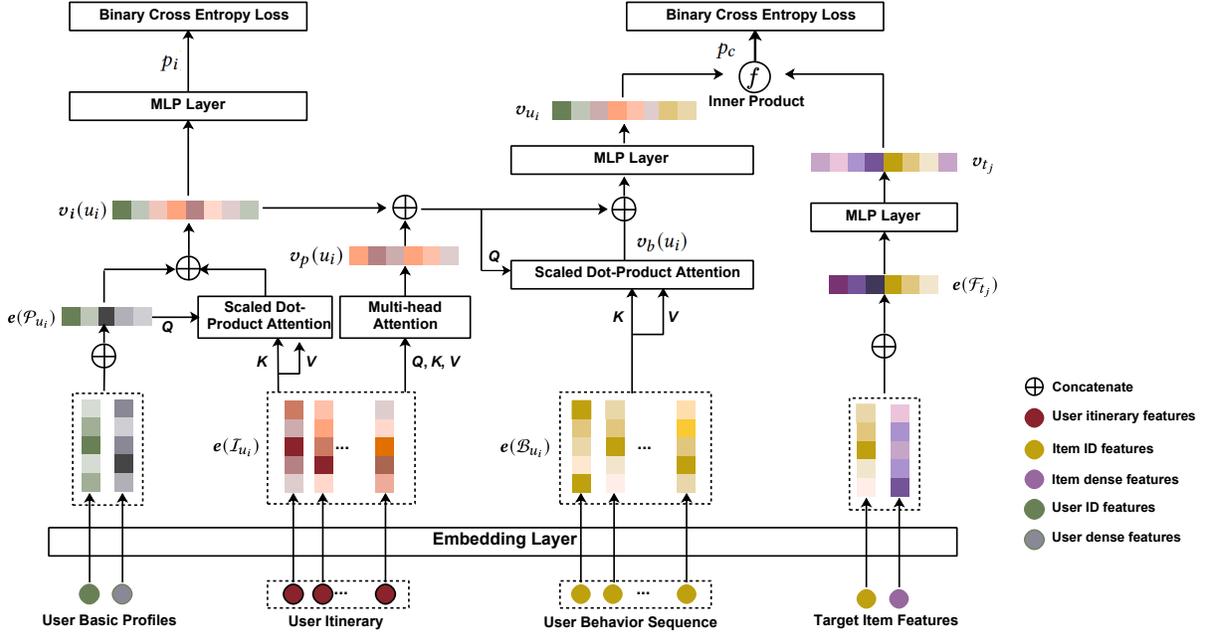}
	\caption{The architecture of the proposed FitNET model.}
	\Description{The structure of the proposed FitNET.}
	\label{fig:2}
\end{figure*}

In the following we present details of the proposed FitNET model for the personalized matching 
task at Fliggy. The architecture of the FitNET is illustrated in Figure \ref{fig:2}. There are four categories of input data in FitNET. It is worth noting that features of input data are either ID features (e.g., "home$\_$city$\_$id" and "item$\_$id") or dense features (e.g., "gender" and "age$\_$level").

\subsection{Feature Representation}
Unlike sponsored search, users come into the recommendation interfaces of FRS without clearly expressing their intentions, which matters for the quality of personalized matching.  
Therefore, features of input data in FRS that depict the intentions of users are the critical elements in FitNET. 
Besides user basic profile and user behavior sequence whose features are frequently used by popular industrial recommendation systems \cite{DIN17, SDM19}, this work proposes to utilize features of a user's itinerary (i.e., features of the user's unconsumed orders) to better capture the user's travel intention. 
According to Definition \ref{DEF-Itinerary}, a user's itinerary is a list of unconsumed orders of the user at Fliggy. Therefore, features of a user's itinerary are actually defined as the features of an order. Similarly, features of a user's behavior sequence including a group of interacted items of the user are defined as the features of an item.  
Table \ref{tab:feaset} summarizes the whole feature set used by FitNET. Features in the feature set come from four categories, namely, user's basic profile $\mathcal{P}_{u_i}$, user's itinerary $\mathcal{I}_{u_i}$, user's behavior sequence $\mathcal{B}_{u_i}$, and target item features $\mathcal{F}_{t_j}$. 
Features belonging to the same category consist of a feature group, and they describe the category from different side.    
In FRS, the value of each feature (e.g, home$\_$city$\_$id="Hangzhou") is transformed into a high-dimensional sparse binary vector (e.g., $[0, \dots, 1, \dots, 0]$ denotes the city Hangzhou in China) by encoding techniques \cite{10.1145/2487575.2488200, 2016Deep}. 
Let $\bm{x}_{ij}$ be the binary encoding vector of the $j$-th feature in the $i$-th feature group, and $k=\sum_n{\bm{x}_{ij}[n]}$. Then, vector $\bm{x}_{ij}$ with $k=1$ refers to a one-hot encoding type, and vector $\bm{x}_{ij}$ with $k>1$ refers to a multi-hot encoding type.
Table \ref{tab:feaset} shows that features of user's itinerary or user's behavior sequence are multi-hot encoding vectors, containing rich information of users' preferences. 
With the sparse binary representations of each feature, we apply a novel deep neural network -- FitNET, to capture the interactions of these features. 

\begin{table*}[htbp]
	\caption{Statistics of features sets used in FitNET. Each feature is encoded into a binary sparse vector.}
	\label{tab:feaset}
	\centering
	\begin{tabular}{lcccc}
		\hline
		Category & Feature Group & Dimensionality & Type & \# Nonzero Ids per Instance \\
		\midrule
		\multirow{5}*{\shortstack{User's Basic Proflile ($\mathcal{P}_{u_i}$)}}&gender&2&one-hot&1\\
		&age\_level& $\sim 10$&one-hot&1\\
		&reside\_city\_id& $\sim 10^3$&one-hot&1\\
		&home\_city\_id& $\sim 10^3$&one-hot&1\\
		&...&...&...&...\\
		\hline
		\multirow{4}*{\shortstack{User's Itinerary ($\mathcal{I}_{u_i}$)}}&ordered\_item\_ids&$\sim 10^{6}$&multi-hot& $\sim 10$\\
		&ordered\_cate\_ids&$\sim 10^{3}$&multi-hot&$\sim 10$\\
		&ordered\_dest\_city\_ids&$\sim 10^{3}$&multi-hot&$\sim 10$\\
		&...&...&...&...\\
		\hline
		\multirow{4}*{\shortstack{User's Behavior \\Sequence ($\mathcal{B}_{u_i}$)}}&clicked\_item\_ids&$\sim 10^{6}$&multi-hot& $\sim 10$\\
		&clicked\_cate\_ids&$\sim 10^{3}$&multi-hot&$\sim 10$\\
		&clicked\_dest\_city\_ids&$\sim 10^{3}$&multi-hot&$\sim 10$\\
		&...&...&...&...\\
		
		\hline
		\multirow{4}*{\shortstack{Target Item Features ($\mathcal{F}_{t_j}$)}}&item\_id&$\sim 10^{6}$&one-hot& 1\\
		&cate\_id&$\sim 10^{3}$&one-hot&1\\
		&dest\_city\_id&$\sim 10^{3}$&one-hot&1\\
		&...&...&...&...\\
		\bottomrule
	\end{tabular}
\end{table*}

\subsection{Embedding Layer}
\label{sec:embed}
As shown in Figure \ref{fig:2}, the input historical data of FitNET are composed of four categories. Features of input data from each category are transformed into feature representations in the form of high-dimensional sparse binary vectors, especially for the representations of ID features. For instance, the number of item ids is about millions at Fliggy, which motivates us to apply the widely-used embedding technique \cite{DBLP:conf/nips/MikolovSCCD13} to embed the feature representations into low-dimensional dense vectors, so as to markedly reduce the cardinality of parameters and facilitate the training and serving of FitNET.  

For the representation vector of the $j$-th feature in the $i$-th feature group (i.e., $\bm{x}_{ij}$), let $\bm{W}^{ij}=[\bm{w}_1^{ij}, \cdots, \bm{w}_n^{ij}, \cdots, \bm{w}_{D_{ij}}^{ij}]\in \mathbb{R}^{D_{ij}\times N_{ij}}$ denote the embedding dictionary for the vector, where $\bm{w}_n^{ij}\in \mathbb{R}^{D_{ij}}$ is an embedding vector with dimensionality of $D_{ij}$, and $N_{ij}$ is the number of distinct values $\bm{x}_{ij}$ might be assigned. We segregate feature representations into two types, whose representations are embedded in different ways.

\begin{itemize}
	\item \textbf{One-hot encoding type.} If the vector $\bm{x}_{ij}$ is one-hot encoding with $\bm{x}_{ij}[n]=1$ , then the embedding representation of $\bm{x}_{ij}$, i.e., $\bm{e}(\bm{x}_{ij})$, is the single embedding vector $\bm{w}_n^{ij}$ in $\bm{W}^{ij}$.   
	\item \textbf{Multi-hot encoding type.} If the vector $\bm{x}_{ij}$ is multi-hot encoding with $\bm{x}_{ij}[n]=1$ and $n\in \{n_1, \cdots, n_{k}\}$, then the embedding representation of $\bm{x}_{ij}$ is the concatenation of a group of embedding vectors in $\bm{W}^{ij}$, represented by $\bm{e}(\bm{x}_{ij})=\{\bm{w}_{n_1}^{ij}, \cdots, \bm{w}_{n_k}^{ij}\}$.   
\end{itemize}

Embedding vectors of features belonging to the same input category are further concatenated during the embedding process. 
After embedding, we obtain one embedding vector $\bm{e}(\mathcal{P}_{u_i})\in \mathbb{R}^{D_p}$ for the input category "user's basic profile" of each $u_i \in \mathcal{U}$, and one embedding vector $\bm{e}(\mathcal{T}_{t_j})\in \mathbb{R}^{D_t}$ for the input category "target item features" of each item $t_j\in \mathcal{P}$. As to the input category "user's itinerary" that consists of a list of user $u_i$'s unconsumed orders, its embedding result is a list of embedding vectors for the unconsumed orders, i.e,  $\bm{e}(\mathcal{I}_{u_i})=\{\bm{e}(\bar{o}_{{u_i},1}), \cdots, \bm{e}(\bar{o}_{{u_i},l}), \cdots, \bm{e}(\bar{o}_{{u_i},|\mathcal{I}_{u_i}|})\}$ with $\bm{e}(\bar{o}_{{u_i},l})\in \mathbb{R}^{D_e}$. The embedding vector of an unconsumed order is the concatenation of embedding vectors of all order features. 
With regard to the input category "user's behavior sequence" that is a sequence of items interacted by user $u_i$, its embedding result is a sequence of embedding vectors for those items, i.e., $\bm{e}(\mathcal{B}_{u_i})=\{\bm{e}(t_{{u_i},1}),\cdots, \bm{e}(t_{{u_i},m}), \cdots, \bm{e}(t_{{u_i},|\mathcal{B}_{u_i}|})\}$ with $\bm{e}(t_{{u_i},m}) \in \mathbb{R}^{D_b}$.
The embedding vector of an interacted item is the concatenation of embedding vectors of all its features. 
Here, $D_*(*\in {p,e,b,t})$ denotes the vector dimension and $|*|$ represents the cardinality of the set $*$. 

\subsection{Itinerary-Aware Attention Mechanisms}
\label{subsec:attention}
Since the information of user itineraries plays an important role in inferring users' travel intentions, a series of itinerary-aware attention mechanisms which learn the interactions between the embedding vectors of users' itineraries and the embedding vectors of other input categories are applied in FitNET. 

\subsubsection{\textbf{Interactions of $\bm{e}(\mathcal{P}_{u_i})$ and $\bm{e}(\mathcal{I}_{u_i})$}}
\label{subsec:p}
In FitNET, the scaled dot-product attention \cite{10.5555/3295222.3295349} is applied to the embedding vectors of users' basic profiles and uses' itineraries (i.e., $\bm{e}(\mathcal{P}_{u_i})$ and $\bm{e}(\mathcal{I}_{u_i})$ for each $u_i\in \mathcal{U}$), to better understand users' travel intentions on the premise of knowing their basic profiles. Take the "home$\_$city$\_$id" feature in user's basic profile for example. If a user's home city is "Hangzhou" while the user's destination city in an unconsumed order of his/her itinerary is also "Hangzhou", then FitNET can infer the user's travel intention as visiting relatives and thus do not recommend local tourist attractions in Hangzhou to the user.
In the attention mechanism, $\bm{e}(\mathcal{P}_{u_i}) \in \mathbb{R}^{D_p}$ is the query vector and $\bm{e}(\mathcal{I}_{u_i})=\{\bm{e}(\bar{o}_{{u_i},l})\}_{l=1}^{|\mathcal{I}_{u_i}|}$ with each $\bm{e}(\bar{o}_{{u_i},l}) \in \mathbb{R}^{D_e}$ is used as both the key vector and the value vector, as shown in Figure \ref{fig:2}. The attention function on the set of queries is 

\begin{equation}
\label{eq:4}
\begin{split}
\text{\textbf{Attention}}(Q, K, V)&=\text{\textbf{Softmax}}\left(\frac{\bm{e}(\mathcal{P}_{u_i})\left[\bm{e}(\mathcal{I}_{u_i})\right]^T}{\sqrt{d}}\right)\bm{e}(\mathcal{I}_{u_i})\\
&= \sum_{l=1}^{|\mathcal{I}_{u_i}|} {\alpha_l \times \bm{e}(\bar{o}_{{u_i},l})},
\end{split}
\end{equation}
where $d$ is a hyperparameter, and 
\begin{displaymath}
\alpha_l=\frac{exp\left(score(\bm{e}(\mathcal{P}_{u_i}), \bm{e}(\bar{o}_{{u_i},l}))\right)}{\sum_{l'=1}^{|\mathcal{I}_{u_i}|} {exp\left(score(\bm{e}(\mathcal{P}_{u_i}), \bm{e}(\bar{o}_{{u_i},l'}))\right)}},
\end{displaymath} 
\begin{displaymath}
score(\bm{e}(\mathcal{P}_{u_i}), \bm{e}(\bar{o}_{{u_i},l})) = \bm{e}(\mathcal{P}_{u_i})\ \bm{W_1}\  \left[\bm{e}(\bar{o}_{{u_i},l})\right]^{T}.
\end{displaymath}

Here, the matrix $\bm{W_1} \in \mathbb{R}^{D_p\times D_e}$ is a learnable parameter in the attention mechanism.

The output vector of this attention mechanism is further concatenated with the embedding vector of user $u_i$'s basic profile to get user $u_i$'s intention vector, denoted by $\bm{v_i}(u_i)$. 

\subsubsection{\textbf{Self-interactions of $\bm{e}(\mathcal{I}_{u_i})$}}
\label{subsec:mh}
Note that a single unconsumed order of a user cannot accurately describe the user's preference for itinerary (such as the preference for destination city or for the item category). Hence, as shown in Figure \ref{fig:2}, a multi-head self-attention mechanism \cite{10.5555/3295222.3295349} is employed in FitNET to capture the self-interactions among embedding vectors of unconsumed orders (i.e., $\{\bm{e}(\bar{o}_{{u_i},l})\}_{l=1}^{|\mathcal{I}_{u_i}|}$) in a user's itinerary, so as to better understand users' purchasing preferences. In the multi-head self-attention mechanism, we employ $N$ heads, where $N$ is set to $8$ by following the recommended setting in \cite{10.5555/3295222.3295349}. The output of each head is concatenated and once again projected using a parameter matrix $\bm{W^O}$, resulting in an output vector $\bm{v}_{p}(u_i) \in \mathbb{R}^{D_m}$ for each user $u_i \in \mathcal{U}$. The attention function is

\begin{equation}
\label{eq:5}
\begin{split}
\text{\textbf{MultiHead}}(Q, K, V) &= \bm{v}_{p}(u_i) \\
&= 	\left[\bm{v}_{p}^1(u_i)\varoplus \cdots \varoplus\bm{v}_{p}^N(u_i)\right]\bm{W^O},
\end{split}
\end{equation}
where $\varoplus$ represents the operation of concatenating two vectors, and the $n$-th head $\bm{v}_{p}^n(u_i) \in \mathbb{R}^{\frac{D_m}{N}}$ in Equation \ref{eq:5} is 
\begin{equation}
\label{eq:6}
\bm{v}_{p}^n(u_i) = \text{\textbf{Attention}}\left(\bm{e}(\mathcal{I}_{u_i})\bm{W_n^Q},\ \  \bm{e}(\mathcal{I}_{u_i})\bm{W_n^K}, \ \ \bm{e}(\mathcal{I}_{u_i})\bm{W_n^V}\right),
\end{equation}
and $\bm{W_n^Q}$, $\bm{W_n^K}$, $\bm{W_n^V} \in \mathbb{R}^{D_e\times \frac{D_m}{N}}$ are the learnable project matrices of the $n$-th head. 

\subsubsection{\textbf{Interactions of $\bm{e}(\mathcal{P}_{u_i})$, $\bm{e}(\mathcal{I}_{u_i})$, and $\bm{e}(\mathcal{B}_{u_i})$}}
\label{subsec:sda2}
As one of important inputs of FitNET, a user's behavior sequence, however, usually contains
causal interactions to items that are unrelated to the user's itinerary.
To filter out the negative impact brought by causal interactions on the acquirement of users' diverse needs, as illustrated in Figure \ref{fig:2}, FitNET adopts another scaled dot-product attention mechanism to emphasize more important interacted items in a user's behavior sequence, according to the user's itinerary information. 
As indicated in the figure, the user intention vector $\bm{v_i}(u_i)$ that encodes both of the basic profile (i.e., $\bm{e}(\mathcal{P}_{u_i})$) and the travel intention information of user $u_i$ is concatenated with the output vector of the multi-head attention mechanism $\bm{v}_{p}(u_i)$ that encodes the preferences of the user, and then the concatenation result (denoted by $\bm{v_c}=\left[\bm{v_i}(u_i)\varoplus \bm{v}_{p}(u_i)\right]$) is used as the query vector in the attention mechanism. The embedding vector of the user $u$'s behavior sequence (i.e., $\bm{e}(\mathcal{B}_{u_i})=\{\bm{e}(t_{{u_i},m})\}_{l=1}^{|\mathcal{B}_{u_i}|}  \text{ with each } \bm{e}(t_{{u_i},m}) \in \mathbb{R}^{D_b}$) is used as both the key vector and the value vector. The attention function on the set of queries is 

\begin{equation}
\label{eq:7}
\begin{split}
\text{\textbf{Attention}}(Q, K, V)&=\text{\textbf{Softmax}}\left(\frac{\bm{v_c}\left[\bm{e}(\mathcal{B}_{u_i})\right]^T}{\sqrt{d}}\right)\bm{e}(\mathcal{B}_{u_i})\\
&= \sum_{m=1}^{|\mathcal{B}_{u_i}|} {\beta_m \times \bm{e}(t_{{u_i},m})},
\end{split}
\end{equation}
where $d$ is a hyperparameter, and the weight $\beta_m$ for the $m$-th interacted item in the user behavior sequence is computed by
\begin{displaymath}
\beta_m=\frac{exp\left(score(\bm{v_c}, \bm{e}(t_{{u_i},m}))\right)}{\sum_{m'=1}^{|\mathcal{B}_{u_i}|} {exp\left(score(\bm{v_c}, \bm{e}(o_{{u_i},m'}))\right)}},
\end{displaymath} 
\begin{displaymath}
score(\bm{v_c}, \bm{e}(t_{{u_i},m})) = \bm{v_c}\ \bm{W_2}\  \left[\bm{e}(t_{{u_i},m})\right]^{T},
\end{displaymath}
where the matrix $\bm{W_2} \in \mathbb{R}^{D_p\times D_b}$ is a learnable parameter in the attention mechanism. The output vector of this attention mechanism is represented as $\bm{v}_(u_i)$ in Figure \ref{fig:2}.

To sum up, by introducing users' itineraries into the attentive procedure over user's behavior sequence, users' diverse needs are more precisely embedded, addressing the diversity problem faced by OTPs.

\subsection{MLP Layer}
The Multi-Layer Perceptron (MLP) layer is a fully-connected feed-forward neural network which employs multiple layers of neurons with nonlinear activation functions, and therefore is more powerful to learn combination of features automatically \cite{Alsmadi2009Performance}. As indicated in Figure \ref{fig:2}, there are totally three MLP layers in FitNET. The first MLP layer utilizes user intention vectors (i.e., $\{\bm{v}_i(u_i)\}$ with each $u_i\in \mathcal{U}$) as the input, and is responsible for learning non-linear relationships of features in these input vectors and finally derives a sightseeing probability by a sigmoid function, denoted by $p_{i}$, representing the travel intention of user $u_i$.  
The input of the second MLP layer is the concatenation of three intermediate result vectors in FitNET, i.e., $[\bm{v}_i(u_i) \varoplus \bm{v}_{p}(u_i) \varoplus \bm{v}_b(u_i)]$. Since the vector $\bm{v}_i(u_i)$ encodes the information of the user's travel intention, the  vector $\bm{v}_{p}(u_i)$ contains the information of the user's preferences, and the vector $\bm{v}_b(u_i)$ encompasses the information of the user's interacted items, the purpose of this MLP layer is to learn a high-quality representation for each user, denoted as $\bm{v}_{u_i}$, by taking those information into account.
Finally, the third MLP layer takes the embedding vectors of target item features as the input, and learns a better representation for each item, denoted as $\bm{v}_{t_j}$. 
After the MLP processing, the learned representations of all users and the learned representations of all target items are executed pairwise calculations according to the inner product operation shown in Equation \ref{eq:3}. The inner product between a user $u_i$'s representation  $\bm{v}_{u_i}$ and a target item $t_j$'s representation $\bm{v}_{t_j}$ is viewed as the probability of the target item $t_j$ being clicked by the user $u_i$ in the recommendation phase, which is denoted by $p_{c}$ in Figure \ref{fig:2}. 

\subsection{Training $\&$ Serving}
In the training phase, the FitNET model is trained according to a loss function that consists of two parts, as shown in Equation \ref{eq:8}. 

\begin{equation}
\label{eq:8}
Loss = Loss_{i} + Loss_{c} 
\end{equation}

The first part in the loss function, i.e., $Loss_{i}$, corresponds to the left '\textit{Binary Cross Entropy Loss}' module in Figure \ref{fig:2}. 
It utilizes the pre-known users' travel intentions with respect to each of their orders as labels to compute the loss caused by predicting users' travel intentions. An indicator variable $I_{i}$ is employed to label each order, with $I_{i}=1$ presenting the current user's intention of creating the current order is for sightseeing, and $0$ otherwise. Specifically, the loss $L_{i}$ is calculated by 

\begin{equation}
\label{eq:9}
Loss_{i} = - I_{i} \log{p_{i}} - (1-I_{i})log(1-p_{i}),
\end{equation}
where $p_{i}$ is the output of the leftmost MLP layer in FitNET, representing the sightseeing probability of a user predicted by FitNET.    

The second part of the loss function, i.e., $Loss_{c}$, corresponds to the right '\textit{Binary Cross Entropy Loss}' module in Figure \ref{fig:2}.
It makes use of users' click behaviors of to recommended items as the labels which are collected before the users consume any order in their itineraries, so as to measure the loss introduced by predicting users' click behaviors to recommended items. Similarly, an indicator variable $I_{c}$ is used to label each <\textit{user-item}> pair, with $I_{c}=1$ indicating the current user will click the current recommended item, and $0$ otherwise. Specifically, the loss $Loss_{c}$ is computed by 

\begin{equation}
\label{eq:10}
Loss_{c} = - I_{c} \log{p_{c}} - (1-I_{c})log(1-p_{c}),
\end{equation}
where $p_{c}$ is the output of the inner product operation in FitNET, representing the click probability of a user to a recommended item predicted by FitNET.   

After training, FitNET can be used as user representation mapping function $f_{\mathcal{U}}$ and item representation mapping function $f_{\mathcal{T}}$. At serving time, the information of a user's basic profile, itinerary, and behavior sequence are fed into the $f_{\mathcal{U}}$ function to derive the representation vector for each user. Meanwhile, the features of a item are input into the $f_{\mathcal{T}}$ function to get the representation vector for every item in the item pool $\mathcal{P}$. Then, these representation vectors are utilized to match top $k$ items from the item pool for each user in terms of the user's travel intention and preference, according to an approximation nearest neighbor search approach \cite{YouTubeDNN16}.
Note that when a user offers new click behaviors, the user's representation vector will accordingly be altered, which means FitNET enables real-time personalization for the matching phase of FRS.

\section{Experiments}
\label{sec:exp}
In this section, we evaluate the effectiveness of the proposed FitNET model extensively. At first, FitNET is compared with several existing methods on an offline real-world dataset. Then, an online A/B test of all comparison models is reported at Fliggy's live production environment. 

\subsection{Comparison Methods}
We compare the proposed FitNET with five related methods.
\begin{itemize}
	\item \textbf{FitNET$^-$}: FitNET$^-$ is a variant of FitNET. 
	Being different from FitNET, only the most recent unconsumed order of a user is used as the input of FitNET$^-$, while FitNET takes all unconsumed orders of a user as a whole (i.e., the user's itinerary)  and fed them to the network.
	\item \textbf{YouTube DNN}\cite{YouTubeDNN16}: YouTube DNN model is able to effectively assimilate many signals and model their interaction with layers of depth, which is one of the most successful deep learning methods used for industrial recommendation systems.
	\item \textbf{DeepWalk}\cite{DeepWalk+18}: DeepWalk is one of the most popular methods in commodity recommendation. It builds an item graph from users' behavior history, and learns the embedding of all items by considering the sequential information of items recorded in the graph. Item embeddings are employed to compute pairwise similarity between any two items, which are then utilized to achieve the recommendation. 
	\item \textbf{MIND}\cite{MIND19}: MIND is deep neural network with behavior-to-interest dynamic routing which deals with user’s diverse interests in the matching stage. MIND is proven to get superior performance than other state-of-the-art methods for recommendation.
	As the interests of users at OTPs are not as divergent as that in other online e-commerce platforms (such as Taobao in China), the number of extracted interest in MIND is set to $1$ (i.e., $k=1$) in the experiment, which has been verified to be optimal setting for the hyperparamter $k$ at Fliggy platform.
	\item \textbf{OrderDest2i}: OrderDest2i is a weak-personalized recommendation method which is a widely used by industrial recommendation systems. OrderDest2i recommends most popular items with respect to the destination city of a user's order, and matched items are further ranked based on their popularity scores.  
\end{itemize}

\subsection{Offline Experiments}

\subsubsection{Dataset}
We employ the Fliggy production dataset in the offline evaluation. The dataset is extracted from \textbf{Mobile Fliggy App}, contains one month users' logs at Fliggy collected in April 2020. 
In the dataset, impression and click samples are labeled as positive samples. On the premise that the number of positive samples is constant, we test the impact of different settings for the amount of negative samples on the performance of FitNET in Section \ref{subsubsec:ns}, and list the best setting in Table \ref{tab:dataset}.
The dataset is further organized into the training dataset and the testing dataset. It is worth noting that an algorithm is designed to automatically label each itinerary of a user in the training dataset with a travel intention, according to a group of well-defined rules set by our experienced staffs. The effectiveness of the algorithm in identifying a user’s travel intention is verified by industrial practices. More details of the Fliggy production dataset are summarized in Table \ref{tab:dataset}. 

\begin{table}[htbp]
	\caption{Statistics of Fliggy production dataset. (M - Million)}
	\label{tab:dataset}
	\centering
	\setlength{\tabcolsep}{3mm}{
		\begin{tabular}{l|cc}
			\hline
			Dataset features	& Training  & Testing \\
			\hline
			\# of samples& 12.1M& 3.4M\\
			\# of positive samples & 1.3M&0.4M \\
			default \# of negative samples&11.8M& 3M \\
			\# of users& 0.27M&  0.1M \\
			\# of items& 1.53M& 1.51M \\
			\# of orders& 0.73M& 0.26M \\
			\# of itineraries& 0.34M& 0.11M \\
			\hline
	\end{tabular}}
\end{table}

\subsubsection{Metrics}
In the offline setting, two metrics, i.e., $\bm{HitRate@k}$ and $\bm{Precision@k}$, are adopted to measure the recommendation performance of different methods, which are also widely used in other related works mentioned in Section \ref{sec:related}. $HitRate@k$ represents the proportion of test cases that have the correctly recommended items in the top $k$ recommendation list of a user, defined as

\begin{equation}
\label{eq:hr}
HitRate@k=\frac{\sum_{(u_i, t_j)\in \mathcal{S}_{test}}I(\text{target item occurs in top-}k)}{| \mathcal{S}_{test}|},
\end{equation}
where $\mathcal{S}_{test}$ denotes the test set consisting of test cases with each test case being in the form of a user-item pair $(u_i,t_j)$, and $I$ is the indicator function. 

$Precision@k$ reflects how many items a user click on within the confines of the top-$k$ recommended items for the user. It is calculated by

\begin{equation}
\label{eq:precision}
Precision@k= \frac{1}{|\mathcal{U}|}\sum_{u_i\in \mathcal{U}}{\frac{|\mathcal{S}_{c}(u_i)@k\cap \mathcal{S}_{gt}(u_i)|}{k}},
\end{equation}
where $|\mathcal{U}|$ is the cardinality of the user set $\mathcal{U}$, $\mathcal{S}_{c}(u_i)@k$ denotes the predicted top-$k$ recommendation items for user $u_i$, and $\mathcal{S}_{gt}(u_i)$ represents the ground truth items that are clicked by user $u_i$.

\subsubsection{Evaluation of Different Settings of Negative Samples}
\label{subsubsec:ns}
With regard to the training of FitNET, positive samples can be directly defined as users' click items with fix cardinality of $1.7$M (see Table \ref{tab:dataset} for details), since clicks indicate users' feedback to a impression item (i.e., recommended item) being a likely match to users' intention. However, defining negative samples for the matching phase in a recommendation system is a non-trivial problem \cite{negativeSampling20, Airbnb18}, for the quality of selected negative samples has great impact on the performance of a recommendation model. For example, a model trained by using the non-click impressions as its negative samples displays the significantly worse performance in retrieving items for users \cite{negativeSampling20}, compared with using random samples as negatives. 
In view of this, before we make the comparison of all methods, we conduct an experiment to evaluate the influence of different settings to negative samples on FitNET.
In particular, we verify three settings of negative samples:

\begin{enumerate}[\textbf{Setting} $1$:]
	\item For each positive sample, we randomly sample items from the item pool as negative samples.
	\item For any user, the user's negative samples are limited to items whose destination city is consistent with that of the user's itinerary. 
	Besides, we ensure that the categories of $50\%$ negative samples are the same as that of their corresponding positive samples and the categories of the remaining $50\%$ negative samples are different from that of their corresponding positive samples. 
	\item The destination city of a user's negative samples is also confined to that of the user's itinerary. And we make sure that only $10\%$ negative samples have the same categories as that of their corresponding positive samples and $90\%$ negative samples have different categories compared with that of their corresponding positive samples.  
\end{enumerate}

The experimental results in terms of the above settings are listed in Tables \ref{tab:str1}-\ref{tab:str3}. We derive three important observations from the three tables:

\begin{enumerate}[$1$)]
	\item FitNET achieves significantly better performance under the settings $2$ and $3$. This is because there is no restriction on the destination city of negatives in setting $1$, and as a result a good many samples whose destination city is irrelevant to the user's itinerary are selected as the negatives. This indicate that confining the destination city of negative samples to that of the user's itinerary is of vital importance in training a travel-related matching model. 
	\item The amount of negative samples has apparent impact on the hit rate and the precision of FitNET. In specific, with the increase of negative samples, two metrics are improved at first, since the increase of negative samples enhances the generalization ability of the trained model in differentiating between positives and negatives. 
	However, when the ratio of negative samples across $10/11$, there is not obvious additional gain over the FitNET model. Therefore, $1$:$10$ is applied as the fix setting to the amount of negative samples in the following experiments.
	\item FitNET gains its best performance under the setting $3$. Thus we can conclude that by augmenting the ratio of negative samples that have different categories compared with that of their corresponding positive items can improve the ability of the model in distinguishing negative items. In the following experiments, the third setting of negative samples is utilized. 
\end{enumerate}

\begin{table*}[htbp]
	\caption{Experiments by varying the amount of negative samples under the setting 1.}\label{tab:str1}
	\centering
	\setlength{\tabcolsep}{1.5mm}{
		\begin{tabular}{l|cccc|cccc}
			\hline
			\multirow{2}*{\shortstack{+:-}} &\multicolumn{4}{c|}{Hit Rate}&\multicolumn{4}{c}{Precision}\\
			\cline{2-9}
			&@3&@10&@20&@50&@3&@10&@20&@50\\
			\hline
			1:1& 0.071& 0.157& 0.23&0.406& 0.029 & 0.025&0.017 &0.011 \\
			1:2& 0.074& 0.163& 0.234&0.413& 0.031&0.028 &0.018 & 0.013\\
			1:5& 0.076& 0.165& 0.239&0.415& 0.034&0.029 &0.020 & 0.015\\  
			1:8& 0.079& 0.167& 0.242&0.418& 0.037&0.031 &0.023 & 0.018\\
			1:10& \textbf{0.08}& \textbf{0.171}&0.245&0.42& 0.041& \textbf{0.032}& \textbf{0.025}&\textbf{0.0186} \\
			1:15& 0.08& 0.169& \textbf{0.246}&\textbf{0.421}& \textbf{0.0413}&0.031& 0.025&0.0185\\  
			\hline 
	\end{tabular}}
\end{table*}

\begin{table*}[htbp]
	\caption{Experiments by varying the amount of negative samples under the setting 2.}\label{tab:str2}
	\centering
	\setlength{\tabcolsep}{1.5mm}{
		\begin{tabular}{l|cccc|cccc}
			\hline
			\multirow{2}*{\shortstack{+:-}} &\multicolumn{4}{c|}{Hit Rate}&\multicolumn{4}{c}{Precision}\\
			\cline{2-9}
			&@3&@10&@20&@50&@3&@10&@20&@50\\
			\hline
			1:1& 0.154& 0.258& 0.357&0.509& 0.058 & 0.0415&0.0296 &0.0212 \\
			1:2& 0.158& 0.263& 0.361&0.517& 0.063&0.0423 &0.0301 & 0.0219\\
			1:5& 0.162& 0.269& 0.365&0.525& 0.0708&0.0436 &0.0309 & 0.0223\\  
			1:8& 0.163& 0.273& 0.37&0.529& 0.0715&0.0448 &0.0312 & 0.0227\\
			1:10& \textbf{0.168}&0.278&\textbf{0.373}&\textbf{0.536}& \textbf{0.0729}& 0.0456& 0.0315&\textbf{0.0235} \\
			1:15& 0.167& \textbf{0.280}& 0.372&0.535& 0.0728&\textbf{0.0457}& \textbf{0.0317}&0.0235\\  
			\hline
	\end{tabular}}
\end{table*}

\begin{table*}[htbp]
	\caption{Experiments by varying the amount of negative samples under the setting 3.}\label{tab:str3}
	\centering
	\setlength{\tabcolsep}{1.5mm}{
		\begin{tabular}{l|cccc|cccc}
			\hline
			\multirow{2}*{\shortstack{+:-}} &\multicolumn{4}{c|}{Hit Rate}&\multicolumn{4}{c}{Precision}\\
			\cline{2-9}
			&@3&@10&@20&@50&@3&@10&@20&@50\\
			\hline
			1:1& 0.16& 0.265& 0.363&0.516& 0.063 & 0.043&0.0303 &0.0221 \\
			
			1:2& 0.163& 0.269& 0.371&0.528& 0.067&0.0438 &0.0307 & 0.0227\\
			1:5& 0.167& 0.274& 0.375&0.533& 0.0714&0.0446 &0.0314 & 0.0231\\  
			1:8& 0.169& 0.281& 0.378&0.536& 0.0725&0.0454 &0.0317 & 0.0235\\
			1:10& \textbf{0.17}&0.286&\textbf{0.38}&0.54& 0.074&\textbf{0.046} & \textbf{0.032}&\textbf{0.024} \\
			1:15& 0.17& \textbf{0.287}& 0.379&\textbf{0.542}&\textbf{ 0.0742}&0.0459& 0.032&0.024\\  
			\hline
	\end{tabular}}
\end{table*}

\subsubsection{Ablation Study}
To evaluate the effectiveness of the proposed three itinerary-aware attention mechanisms in the FitNET model (see Section \ref{subsec:attention} for details), we conducted an ablation study by comparing FitNET with its three variants, named as $Var_1$, $Var_2$, and $Var_3$. $Var_1$ removes the attention mechanism proposed in Section \ref{subsec:p} from FitNET, which captures the interactions between users' itineraries and their basic profiles. Note that, since the removed attention mechanism is highly related to the auxiliary tower in the FitNET architecture, corresponding to the $Loss_i$ part in the loss function of FitNET (see Equation \ref{eq:8}) which is set to minimize the loss of predicting users' travel intentions, the auxiliary tower is deleted from the architecture of FitNET in $Var_1$. $Var_2$ removes the multi-head attention mechanism proposed in Section \ref{subsec:mh} from FitNET, which learns the self-interactions among embedding vectors of unconsumed orders in a user's itinerary. Finally,  $Var_3$ cuts out the attention mechanism proposed in Section \ref{subsec:sda2} from FitNET, which emphasizes more important interacted items in a user's behavior sequence, according to the interactions among the user's itinerary, profile, and behavior sequence. Experimental results of the ablation study are listed in Table \ref{tab:ablation}. Table \ref{tab:ablation} shows that: 
\begin{enumerate}[$1$)]
	\item Each of the three attention mechanisms makes considerably contribution to improve the hit rate and precision, due to the concern of interactions between itineraries and other input categories, which confirms again the important role of itineraries in improving the matching phase at OTPs. 
	\item Without the attention mechanism proposed in Section \ref{subsec:p} and the auxiliary tower that minimizes the loss in identifying a user’s travel intention, there is a relatively bigger decrease in hit rate and precision, which on one hand demonstrates the importance of retrieving items for users in terms of their travel intentions and on the other the effectiveness of using the itinerary in referring travel intentions. 
\end{enumerate}

\begin{table*}[htbp]
	\caption{Ablation Study.}\label{tab:ablation}
	\centering
	\setlength{\tabcolsep}{1.5mm}{
		\begin{tabular}{c|cccc|cccc}
			\hline
			\multirow{2}*{\shortstack{Methods}} &\multicolumn{4}{c|}{Hit Rate}&\multicolumn{4}{c}{Precision}\\
			\cline{2-9}
			&@3&@10&@20&@50&@3&@10&@20&@50\\
			\hline
			FitNET & \textbf{0.17} & \textbf{0.286} & \textbf{0.38} & \textbf{0.5403} & \textbf{0.074} & \textbf{0.046} & \textbf{0.032} & \textbf{0.024} \\
			Var$_1$ & 0.162 & 0.269 & 0.368 & 0.5312 & 0.0713 & 0.0441 & 0.031 & 0.023 \\
			Var$_2$ & 0.164 & 0.267 & 0.372 & 0.5346 & 0.0718 & 0.044 & 0.0312 & 0.0236 \\  
			Var$_3$ & 0.168 & 0.273 & 0.376 & 0.535 & 0.0723 & 0.0452 & 0.0317 & 0.0238 \\
			\hline
	\end{tabular}}
\end{table*}

\subsubsection{Evaluation of Comparison Methods}
Results on the offline dataset of different methods are shown in Table \ref{tab:offtotal}. 
In general, OrderDest2i which executes personalized matching simply based on the destination city of the order is beaten by all deep learning based methods (i.e., FitNET, FitNET$^-$, MIND, YouTube DNN, and DeepWalk), revealing the power of deep learning for improving the matching phase of recommendation systems.
It can also be observed that methods aware of the information of user's itinerary (i.e., FitNET and FitNET$^-$) apparently outperform other methods. This confirms that the consideration of user's itinerary in the design of a deep matching network is of great help in solving the three challenges, namely, sparsity, diversity, and implicitness, facing OTPs. 
Another important observation from the table is that FitNET gains higher hit rate and precision compared with its variant, FitNET$^-$. Restate that FitNET$^-$ trains the network only based on the most recent unconsumed order of a user, while all unconsumed orders in a user's itinerary are used in FitNET. According to Fliggy's experience, a user's travel intention with respect to the most recent order is probably returning home rather than going sightseeing, and users usually show very low interest in the recommended items relating to their homes. Therefore, compared with FitNET$^-$, FitNET gets on average $1.4\%$ improvement in terms of hit rates, and on average $1.7\%$ growth in terms of precision. This observation once again confirms that all unconsumed orders of a user must be treated as a whole to elevate the prediction accuracy of the user's travel intention.  
The table also demonstrates that MIND beats YouTube DNN due to the usage of label-aware attention mechanism over the embedding of users' behavior sequences, which decreases the importance of irrelevant interacted items of a user and as a result better identifies a user's diverse interests. On the other hand, YouTube DNN simply uses the average pooling strategy over the embeddings of users' behavior sequences, reducing the quality of its matched items.   
Finally, it is worth noting that FitNET makes on average $5.2\%$ increase in hit rate and on average $9.1\%$ improvement in precision, compared with the next best method, MIND. Because, FitNET is able to learn more accurate information of users' travel intentions and preferences, by being awareness of users' itinerary information in the design of the matching framework for OTPs.

\begin{table*}[htbp]
	\caption{Comparison of different methods on the offline dataset.}\label{tab:offtotal}
	\centering
	\setlength{\tabcolsep}{1.5mm}{
		\begin{tabular}{l|cccc|cccc}
			\hline
			\multirow{2}*{\shortstack{Methods}} &\multicolumn{4}{c|}{Hit Rate}&\multicolumn{4}{c}{Precision}\\
			\cline{2-9}
			&@3&@10&@20&@50&@3&@10&@20&@50\\
			\hline
			FitNET& \textbf{0.17}& \textbf{0.286}& \textbf{0.38}&\textbf{0.54}&\textbf{0.074} & \textbf{0.046}&\textbf{0.032} &\textbf{0.024} \\
			$\mathrm{FitNET}^{-}$& 0.166& 0.282& 0.376&0.537& 0.0732&0.0454 &0.0313 & 0.0235\\
			MIND& 0.156& 0.271& 0.365&0.528& 0.062&0.042 &0.031 & 0.023\\  
			YouTube DNN& 0.13& 0.25& 0.34&0.48& 0.051&0.039 &0.028 & 0.021\\
			DeepWalk& 0.09& 0.2&0.29&0.43& 0.047& 0.035& 0.023&0.016 \\
			OrderDest2i& 0.045& 0.14& 0.19&0.35& 0.033&0.024& 0.016&0.009\\  
			\hline
	\end{tabular}}
\end{table*}

\subsection{Online A/B Test}
We conduct online experiments by deploying FitNET to handle real traffic in the personalized interfaces of Mobile Fliggy App (see Figure \ref{fig:interface}) for seven days in May 2020. We deploy all comparison methods concurrently in the matching phase of Fliggy platform under an A/B test framework, and one thousand of candidate items are retrieved by each method in the matching phase, which then fed to the ranking phase. To guarantee the fairness of comparisons, retrieved items by each method are ranked by the same ranking engine. 
Moreover, the scheduling engine is revised to make sure the personalized interfaces of each method obtain $20\%$ daily traffic at Fliggy, to make sure that each method has equal traffic.
\textit{\textbf{click-through rate}} ($\bm{CTR}$ for short), a widely-used industrial metric \cite{DeepWalk+18, MIND19, SDM19}, is employed to evaluate the performance of recommendation methods for serving online traffic. 
CTR is calculated by taking the total number of times items are clicked on and dividing it by the measured item impressions, as shown in Equation \ref{eq:ctr}.
\begin{equation}
\label{eq:ctr}
CTR=\frac{\# \text{ of clicks}}{\#\text{ of impressions}}
\end{equation}

\begin{figure}[h]
	\centering
	\includegraphics[width=\linewidth]{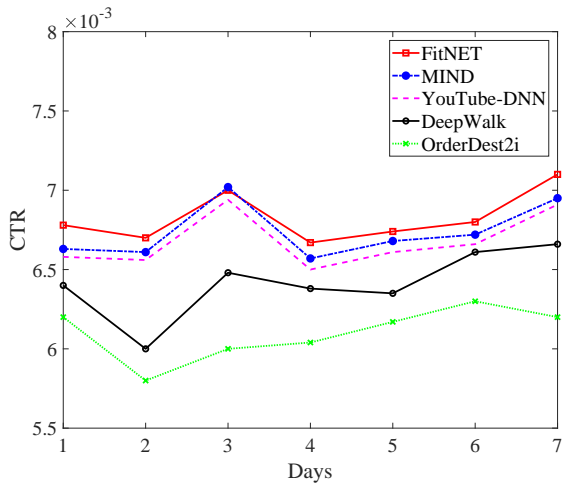}
	\caption{Online CTRs of different methods in seven days in May 2020.}
	\label{fig:online}
\end{figure}

The experimental results are shown in Figure \ref{fig:online}. It is clear that FitNET outperforms other recommendation methods, which indicates that FitNET generates better representations for users by taking their itinerary information into account. 
Besides, we make several observations from the figure. First, OrderDest2i exhibits the worst performance as it recommends items to users simply based on the information of destination city. This once again confirm our argument that a user's travel intention or preference can rarely be captured by the information in a single order. Second, MIND which models diverse interests of users achieves competitively CTR, compared with other baseline methods. This indicates that the acquirement of users' diverse needs is vital for enhancing the personalization of the matching phase. As to FitNET, it not only catches the diversity of users' needs by proposing an attentive learning mechanism based on a user's behavior sequence (see Section \ref{subsec:sda2} for details), but also makes use of the important itinerary information of users to elevate the quality of personalized matching results.  
We are delighted to observe that the average CTR of FitNET is $1.3\%$ higher than that of the next best method, i.e., MIND, which is an apparent progress made by a industrial recommendation system.

\section{Conclusion and Future Work}
Fliggy, one of the most popular online travel platforms (OTPs) in China, has urgent need in helping users find travel-related items they might need or prefer from a item pool with millions of cardinality. Although the popular deep matching networks which retrieve personalized items for users based on deep neural networks achieve many successes in recent years, tremendous stress still has been put on the recommendation systems of OTPs, in terms of sparsity, diversity and implicitness.
To address these challenges, we present a novel Fliggy itinerary-aware deep matching network (FitNET) in this paper, which focuses on improving the accuracy of retrieved items for a user in the matching phase. 
In particular, to cope with the implicitness problem, we introduce the concept of \textit{itinerary} that contains all unconsumed orders of a user, and uses all orders in a user's itinerary rather than a single order to improve the prediction accuracy to the user's travel intention.
Then, to address the diversity and cold start problems, users' profiles are incorporated into the FitNET model.
Meanwhile, a group of itinerary-aware attention mechanisms are proposed in FitNET, which are very helpful not only in inferring users' travel intentions or preferences, but also in distinguishing users' diverse needs. 
Offline experiments are conducted to verify the effectiveness of FitNET in improving recommending accuracy. 
Online CTRs are also reported to demonstrate the effectiveness and feasibility of FitNET in Fliggy's live production environment. 
Experimental results show that FitNET achieves superior performance than state-of-the-art methods for recommendation. It is the consideration of users' itinerary information that enables it retrieves more accurate items for each user.

For future work, we will pursue two directions. 
The first is to incorporate users' comments towards items at Fliggy into our model, 
so as to achieve a better user representations.
The second direction is to upgrade the proposed model by handling more complex scenarios, in which a user may own more than two itineraries.

\begin{acks}
We would like to thank colleagues of our team - Quan Lu, Zhi Jiang, Qi Rao, Wei Wang, Xiaolei Zhou, Shihao Li and Fei Xiong for useful discussions and supports on this work. We also thank the anonymous reviewers for their valuable comments and suggestions that help improve the quality of this manuscript. This work is supported by the National Natural Science Foundation of China (Nos. 62067001 and 62062008); Special funds for Guangxi BaGui Scholars; and is partially supported by the Guangxi Natural Science Foundation (Nos.  2019JJA170045 and 2018JJA170194).
\end{acks}

\bibliographystyle{ACM-Reference-Format}
\bibliography{main}

\end{document}